\documentclass[prb,twocolumn]{revtex4}

\usepackage{graphicx}
\usepackage{amsfonts}
\usepackage{bm}

\newcommand{\vnabla}{{\mbox{\boldmath$\nabla$}}}

\newcommand{\vS}{{\mbox{\boldmath$S$}}}

\newcommand{\vk}{{\mbox{\boldmath$k$}}}
\newcommand{\ver}{{\mbox{\boldmath$r$}}}

\newcommand{\vtau}{{\mbox{\boldmath$\tau$}}}

\newcommand{\tomega}{\tilde{\omega}}
\newcommand{\tDelta}{\tilde{\Delta}}
\newcommand{\tpsi}{\tilde{\psi}}

\newcommand{\valpha}{{\mbox{\boldmath$\alpha$}}}
\newcommand{\vsk}{{\small \mbox{\boldmath$k$}}}

\newcommand{\vg}{\mbox{\boldmath$g$}}
\newcommand{\vq}{\mbox{\boldmath$q$}}
\newcommand{\vD}{\mbox{\boldmath$D$}}
\newcommand{\vDelta}{\mbox{\boldmath$\Delta$}}

\newcommand{\hve}{\hat{\mbox{\boldmath$e$}}}

\newcommand{\hvg}{\hat{\mbox{\boldmath$g$}}}

\newcommand{\vsig}{\mbox{\boldmath$\sigma$}}

\begin{document}
\title{Superconductors without an inversion center of symmetry: The s-wave state. }

\author{P.A. Frigeri$^1$, D.F. Agterberg$^2$, I. Milat$^1$ and
  M. Sigrist$^1$}
\address{$^1$Theoretische Physik ETH-H\"onggerberg CH-8093 Z\"urich,
  Switzerland}
\address{$^2$Department of Physics, University of Wisconsin-Milwaukee,
  Milwaukee, WI 53201}


\begin{abstract}
In materials without an inversion center of symmetry the spin
degeneracy of the conducting band is lifted by an antisymmetric
spin orbit coupling (ASOC). Under such circumstances, spin and
parity cannot be separately used to classify the Cooper pairing
states. Consequently, the superconducting order parameter is
generally a mixture of spin singlet and triplet pairing states. In
this paper we investigate the structure of the order parameter and
its response to disorder for the most symmetric pairing state
($A_1$). Using the example of the heavy Fermion superconductor
CePt$_3$Si, we determine characteristic properties of the
superconducting instability as a function of (non-magnetic)
impurity concentrations. Moreover, we explore the possibility of
the presence of accidental line nodes in the quasiparticle gap.
Such nodes would be essential to explain recent low-temperature
data of thermodynamic quantities such as the NMR-$T_1^{-1}$,
London penetration depth, and heat conductance.
\end{abstract}


\maketitle

\section{ Introduction}

Early studies of superconductivity in materials without inversion
symmetry addressed two-dimensional systems, such as thin films,
interfaces, and surfaces \cite{Edelstein1989, Gorkov2001,
Barzykin2002,Dimitrova2003}. The recent discovery of the
non-centrosymmetric heavy Fermion superconductor CePt$_3$Si
\cite{Bauer2004} has drawn attention towards bulk materials.
Particular interest arose with the observation that the upper
critical field of CePt$_3$Si exceeds the paramagnetic limiting
field considerably. The simplest interpretation of this finding is
in terms of spin triplet pairing.  However, this is in conflict
with the common believe that the absence of an inversion center
prevents electrons from forming spin-triplet pairs
 \cite{Anderson1984}. Indeed, inversion together with time reversal
 are the key symmetries required for Cooper pairing.
In a time reversal invariant system, the lack of inversion
symmetry is connected with the presence of an antisymmetric
spin-orbit coupling (ASOC). This can be represented in the
single-particle Hamiltonian by a term of the general form
\begin{equation}
\alpha {\bm g}_{\bm k} \cdot {\bm S}
\end{equation}
where the vector function $ {\bm g}_{\bm k} $ is odd in $ {\bm k}
$ ($ {\bm g}_{\bm k} = - {\bm g}_{- {\bm k}} $) and $ \alpha $
denotes the coupling strength \cite{Frigeri2004}. The ASOC is
indeed detrimental for most spin triplet pairing states as noted
by Anderson \cite{Anderson1984}. However, it was also found that
triplet states  whose $ {\bm d} $-vector lies parallel to $ {\bm
g}_{\bm k} $ would nevertheless be stable. Such a spin triplet
state has the full symmetry of the crystal point group, we will
therefore call it the $ S $-triplet state from now on.

The interpretation of the absence of paramagnetism in terms of
spin triplet pairing is not unique. Paramagnetic limiting is also
drastically reduced for spin singlet states in the presence of
ASOC
\cite{Bulaevskii1975,Frigeri2004,Frigeri22004,Samokhin2005,Mineev2005,Kaur2005}.
In fact, the absence of the inversion symmetry leads to a
breakdown of the strict classification into even-parity
spin-singlet and odd-parity spin-triplet pairing; these states are
mixed, resulting in a state containing both components
\cite{Edelstein1989, Gorkov2001, Barzykin2002}. Since all but one
spin-triplet states are suppressed by strong enough ASOC (i.e. $
\alpha \gg k_B T_c $, which is usually the case), we concentrate
here on the stable $ S $-triplet state. This ''high-symmetry''
state mixes with the ''$s$-wave'' spin-singlet state (which also
has the full symmetry of the crystal point group), since both of
them belong to the same trivial $A_1$  representation of $C_{4v}
$, the generating point group in CePt$_3$Si. We will call this
combined phase the  ''$s$-wave state''.

The presence of ASOC leads to a splitting of the electron bands by
lifting the spin degeneracy. Thus, the discussion of the
superconductivity is in some sense a two-band problem in this
case. Assuming that superconductivity is restricted to a single
band the
 basis function for irreducible representations of point group C$_{4v}$ have already  been
determined by various groups \cite{Samokhin2004, Sergienko2004,
Mineev2004}. These studies  found that that the quasiparticle gap
for the (most symmetric) $ A_1 $ state would have the form $\Delta
\propto k^{2}_{x}+k^{2}_{y}+ck^{2}_{z}$ which is nodeless in
general. In contrast to these earlier works, we will examine here
the full two-band situation. We show how the symmetry properties
of the pairing interaction and the distribution of the density of
states on the two Fermi surfaces can influence the form of the
pairing state and, in particular, can introduce (accidental) line
nodes in the quasiparticle gap which are not dictated by symmetry.
Since these nodes are accidental their position in momentum space
will vary both with temperature and magnetic fields. It is also
interesting to see how the $s$-wave state is affected by
non-magnetic impurities. These properties could be important to
obtain information on the complex structure of the pairing state
in this material.

\section{Model with antisymmetric spin-orbit coupling}
%

The basic model used to describe the conduction electrons in
crystals without an inversion center can be written as
\begin{equation}
{\cal H}_0 = \sum_{\vsk,s,s'} \left[ \xi_{\vsk} \sigma_0 + \alpha
  \vg_{\vsk} \cdot \vsig \right]_{ss'} c_{\vsk s}^{\dag} c_{\vsk s'},
\label{eq-1}
\end{equation}
where $ c_{\vsk s}^{\dag} $ ($ c_{\vsk s} $) creates (annihilates)
an electron with  wave vector $ \vk $  and spin $s$,
$\hat{\vsig}=(\hat{\sigma}_{x}, \hat{\sigma}_{y},
\hat{\sigma}_{z})$ is the vector of Pauli matrices and
$\hat{\sigma}_{0}$ is the unit matrix  \cite{Frigeri2004}. The
band energy $ \xi_{\vsk} = \epsilon_{\vsk} - \mu $ is measured
relative to the chemical potential $ \mu $. The antisymmetric
spin-orbit coupling (ASOC) term $\alpha \vg_{\vsk} \cdot \vsig$ is
different from zero only for crystals without an inversion center
and can be derived microscopically by considering the relativistic
corrections to the interaction of the electrons with the ionic
potential \cite{Dresselhaus1955,Frigeri22004}. For qualitative
studies, it is sufficient to deduce the structure of the
$\vg$-vector from symmetry arguments \cite{Frigeri2004} and to
treat $\alpha$ as a parameter. We set $ \langle \vg_{\vsk}^2
\rangle = 1 $, where $\langle \rangle$ denotes the average over
the Fermi surface. The ASOC term lifts the spin degeneracy by
generating two bands with different spin structure. The normal
state Green's function,

\begin{equation}
\hat{G}_0(\vk, i \omega_n) = G^{0}_{+} (\vk, i \omega_n)
\hat{\sigma}_{0} +
  (\hvg_{\vsk} \cdot  \hat{\vsig}) G^{0}_{-}(\vk, i \omega_n),
\end{equation}
with
\begin{equation}
G^{0}_{\pm} (\vk, i \omega_n) = \frac{1}{2} \left[ \frac{1}{i
\omega_n -
    \xi_{\vsk} - \alpha |\vg_{\vsk}|} \pm \frac{1}{i \omega_n -
\xi_{\vsk} + \alpha | \vg_{\vsk}|} \right],
\end{equation}
and  $ \hvg_{\vsk} = \vg_{\vsk} / | \vg_{\vsk} | $ ($ | \vg | =
\sqrt{\vg^2} $),
can be diagonalized into the components corresponding to the two
bands, using the unitary transformation
\begin{equation}
\label{Utrans} \hat{U} (\vsk)  = \cos(\theta_{\vsk}/2)-i
\sin(\theta_{\vsk}/2)(\cos \phi_{\vsk} \hat{\sigma}_{y}-\sin
\phi_{\vsk} \hat{\sigma}_{x}),
\end{equation}
where $\vg_{\vsk}=|\vg_{\vsk}|(\sin \theta_{\vsk} \cos
\phi_{\vsk}, \sin \theta_{\vsk} \sin \phi_{\vsk}, \cos
\theta_{\vsk})$ defines the angles $\theta_{\vk}$ and
$\phi_{\vk}$. This allows us to express,
\begin{equation}
\label{Green0band} \hat{G}_{0} (\vk, i \omega_n) = G^{0}_{1} (\vk,
i \omega_n) \hat{\sigma}_{1}(\vk)+G^{0}_{2} (\vk, i \omega_n)
\hat{\sigma}_{2}(\vk),
\end{equation}
through the two Green's functions,
\begin{equation}
G^{0}_{1,2} (\vk, i \omega_n) = \frac{1}{i \omega_n -
    \xi_{1,2}(\vsk)},
\end{equation}
where the  quasi-particle bands $\xi_{1,2}(\vsk) \equiv \xi_{\vsk}
\pm \alpha |\vg_{\vsk}|$ are split by the presence of the ASOC.
The spin structure in the bands is described by the 2 x 2 matrix
\begin{eqnarray}
\sigma^{1}_{\lambda, \mu}(\vk) &\equiv& U_{\lambda, 1}(\vsk)U^{\dag}_{1,\mu } (\vsk)=1/2(\sigma_{0}+(\hvg_{\vsk} \cdot \vsig))_{\lambda, \mu}\nonumber \\
\sigma^{2}_{\lambda, \mu}(\vk) &\equiv& U_{\lambda, 2}(\vsk)U^{\dag}_{2,\mu } (\vsk)=1/2(\sigma_{0}-(\hvg_{\vsk} \cdot \vsig))_{\lambda, \mu}. \nonumber \\
\end{eqnarray}
which act as projection operators in spin space. Since the spins
of quasiparticles in the two bands have opposite directions for a
given momentum $\vk $, the matrices satisfy the conditions,
\begin{eqnarray}
\hat{\sigma}_{1}(\vk) \hat{\sigma}_{2}(\vk)&=&0 \nonumber \\
\hat{\sigma}^{2}_{1,2}(\vk)&=&\hat{\sigma}_{1,2}(\vk)
\end{eqnarray}
and are reciprocal under the inversion operation,
\begin{equation}
\hat{\sigma}_{1,2}(-\vk)=\hat{\sigma}_{2,1}(\vk).
\end{equation}

\begin{figure}[h]
\begin{center}
    \includegraphics[width=8.5cm, height=4cm ]{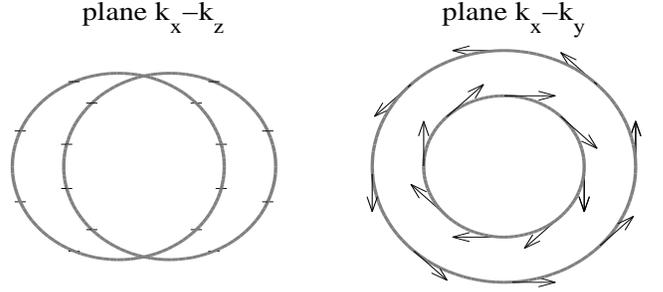}
\begin{minipage}[c]{8cm}
\caption{\label{Fermi} \small{Fermi surfaces for $\vg_{\vsk}
\propto (k_{y}, -k_{x}, 0)$ as in CePt$_{3}$Si. The arrows show
the structure of the quasi-particle spin. Only along the z-axis
the spin degeneracy is preserved.}}
\end{minipage}
\end{center}
\end{figure}

Symmetry considerations lead to the following form for $ {\bm
g}_{\bm k} $ in CePt$_{3}$Si in lowest order expansion in the wave
vector $ {\bm k} $:  $\vg_{\vsk} \propto (k_{y}, -k_{x}, 0)$. The
resulting spin structure is visualized in Fig.\ref{Fermi}, where
we assume for simplicity that the original Fermi surface is
spherical.

The pairing interaction is generally given by
\begin{equation}
{\cal H}_{pair} = \frac{1}{2} \sum_{\vsk,
\vsk'}\sum_{s,s'}V_{s_{1}s_{2},s'_{2}s'_{1}}(\vsk, \vsk')
c^{\dag}_{\vsk s_{1}} c^{\dag}_{-\vsk s_{2}} c_{- \vsk' s'_{2}}
c_{\vsk' s'_{1}} \; ,
\end{equation}
where  $V_{s_{1}s_{2},s'_{2}s'_{1}}(\vsk, \vsk')$ is the pairing
potential with the symmetry properties,
\begin{eqnarray}
 &&V_{s_{1}s_{2},s'_{2}s'_{1}}(\vsk, \vsk')=-V_{s_{2}s_{1},s'_{2}s'_{1}}(-\vsk, \vsk') \nonumber \\
 &=&-V_{s_{1}s_{2},s'_{1}s'_{2}}(\vsk, -\vsk')=V^{*}_{s'_{1}s'_{2},s_{2}s_{1}}(\vsk', \vsk).
\end{eqnarray}

In the absence of the ASOC the system is inversion symmetric in
$\vk$-space and  the interaction factorizes in an orbital and a
spin part. The conventional s-wave superconducting state belongs
to the trivial representation $A_{1g}$ of the inversion symmetric
point group $G$. When the ASOC is turned on,  the point group is
reduced to the non-inversion symmetric subgroup $G'$, whose
trivial  representation $A_{1}$ does not have a definite parity.
In fact $A_{1}$ is compatible with both odd and even
representations of $ G$, i.e. with both $\{A_{1g}, \Gamma_{u}
\}$, where $\Gamma_{u}$ is an odd parity irreducible
representation in $G$,for which $\vg_{\vsk} $ is a basis state
\cite{footnote}. Hence, the general form of the pairing potential
relevant for the realization of the superconducting state
belonging to $ A_1 $ of $G'$ has to involve both basis functions.
Explicitly written, this leads to the form
\begin{eqnarray}
\label{Inter}
&& V_{s_{1}s_{2},s'_{2}s'_{1}}(\vsk, \vsk')= \frac{V}{2} \left\{ e_{s} \; \hat{\tau}_{s_{1}s_{2}} \hat{\tau}^{\dag}_{s'_{2}s'_{1}} \nonumber \right. \\
&+&  e_{t} \left[(\vg_{\vsk}\cdot \hat{\vtau})_{s_{1}s_{2}} (\vg_{\vsk'}\cdot \hat{\vtau})^{\dag}_{s'_{2}s'_{1}}\right]   \nonumber \\
&+& \left. e_{m} \left[( \vg_{\vsk}\cdot
\hat{\vtau})_{s_{1}s_{2}}\hat{\tau}^{\dag}_{s'_{2}s'_{1}}+\hat{\tau}_{s_{1}s_{2}}(\vg_{\vsk'}\cdot
\hat{\vtau})^{\dag}_{s'_{2}s'_{1}} \right]\right\} ,
\end{eqnarray}
where $\hat{\tau}_{s_{1}s_{2}}=( i
\hat{\sigma}^{y})_{s_{1}s_{2}}$, and, $\hat{\vtau}_{s_{1}s_{2}}=(
i \vsig  \hat{\sigma}^{y})_{s_{1}s_{2}}$. To avoid ambiguity we
set $V>0$, and $e_{s}^{2}+e_{t}^{2}+e_{m}^{2}=1$.

The first term of Eqn. (\ref{Inter}) is diagonal in the
conventional s-wave pairing channel, the second one for the
$S$-triplet pairing channel \cite{Frigeri2004}. The last term
describes the scattering of Cooper pairs between the two channels,
which is a result of the absence of inversion symmetry. The
microscopic origin of this last term is explained in the Appendix
\ref{em}. In particular we show the existence of
Dzyaloshinskii-Moriya \cite{Dzyaloshinskii1958, Moriya1960}  type
of interaction for both a weakly interacting Fermi liquid and for
a Hubbard model near half filling. This kind of interaction gives
a contribution to $e_m$.  This can give rise to a large
interaction when the derivative in the density of states is large.
This is the case close to a Van Hove singularity.

\section{Impurity scattering}

Effects of disorder are described by potential scattering of the
quasiparticles, which in real-space representation is given by
\begin{eqnarray}
{\cal H}_{imp} = \sum_{i} {\cal H}_{i}, \quad {\cal H}_{i}=\int
u(\ver-\ver_{i}) \psi^{\dag}_{s}(\ver) \psi_{s}(\ver) d\ver,
\end{eqnarray}
where $u(\ver)$ is the potential of a non-magnetic impurity, which
we consider rather short-ranged such that $s$-wave scattering is
dominant. We are interested in the disorder-averaged normal and
anomalous Green's functions $\hat{G}$ and $\hat{F}$,
\begin{eqnarray}
    \label{green_function}
    G_{\lambda \mu}(r-r')&=&- \langle T_{\tau} \{
    \psi_{\lambda}(r) \psi^{\dag}_{\mu}(r')\} \rangle \nonumber \\
    F_{\lambda \mu}(r-r')&=& \langle T_{\tau} \{
    \psi_{\lambda}(r) \psi_{\mu}(r')\} \rangle \nonumber \\
    F^{\dag}_{\lambda \mu}(r-r' )&=& \langle T_{\tau} \{
    \psi^{\dag}_{\lambda}(r) \psi^{\dag}_{\mu}(r') \}
    \rangle,
\end{eqnarray}
with $r=(\ver,\tau)$ and where the bracket denotes the thermal
average.

For the impurity average we use the Born approximation
\cite{Abrikosov1975},  neglecting
 the possibility of more than two scattering events at the same impurity, which is legitimate if the potential is small in comparison with the characteristic electron energy scale $\epsilon_{F}$ ($\epsilon_{F}$: Fermi energy, or analogue to the band width).

\begin{figure}[h]
\begin{center}
 \includegraphics[width=5cm, height=2.4cm ]{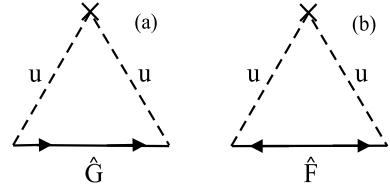}
\begin{minipage}[c]{8cm}
\caption{\label{SelfEnergy} \small{The self energies contribution
due to the impurities scattering in the Born approximation, of
normal type $\Sigma_{G}$ (a) and of anomalous type $\Sigma_{F}$
(b).}}
\end{minipage}
\end{center}
\end{figure}

Formally, the impurity scattering enters the self-energy of the
Greens function of the normal,
 $\hat{\Sigma}_{G}$, Fig. \ref{SelfEnergy}(a), and the anomalous type, $\hat{\Sigma}_{F}$, Fig. \ref{SelfEnergy}(b). Their mathematical expressions read
\begin{eqnarray}
       \label{SelfEnergyEqn}
    \hat{\Sigma}_{G}(i \omega_{n})
    &=&\frac{\Gamma}{\pi N_{0}} \int  \frac{d\vk'}{(2\pi)^{3}} \hat{G}(\vk',i \omega_{n})
    \nonumber \\
    \hat{\Sigma}_{F}(i \omega_{n})&=&\frac{\Gamma}{\pi N_{0}} \int \frac{d\vk'}{(2\pi)^{3}}  \hat{F}(\vk', i \omega_{n}),
\end{eqnarray}
where $\Gamma \equiv \pi n_{imp} N_{0} u^{2}$ is the averaged
scattering rate, $N_{0}\equiv (N_{1}+N_{2})/2$ and $N_{1,2}$ are
the densities of state (DOS) of the two bands at the Fermi level.
We introduced the impurity concentration $n_{i}$ and the $s$-wave
scattering potential $u^{2}$.

The Gor'kov equations with these self-energy contributions are
formally analogous to those obtained for system with an inversion
symmetry \cite{Abrikosov1975}
\begin{eqnarray}
    \label{Gorkov1}
    \left(\hat{G}^{-1}_0(\vk,i\omega_n)-\hat{\Sigma}_{G}(i\omega_{n}) \right) \hat{G}(\vk,i\omega_n) \nonumber \\
    \quad \quad +\left(\hat{\Delta}(\vk)+\hat{\Sigma}_{F}(i\omega_{n})\right) \hat{F}^{{\dag}}(\vk,i\omega_n)&=&\hat{\sigma}_0 \nonumber \\
    \left(\hat{G}^{{-1}^{\top}}_0(-\vk,-i\omega_n)+\hat{\Sigma}^{\top}_{G}(-i\omega_{n})\right) \hat{F}^{{\dag}}(\vk,i\omega_n) \nonumber \\
    \quad \quad
    -\left(\hat{\Delta}^{\dag}(\vk)+\hat{\Sigma}^{\dag}_{F}(i\omega_{n})\right) \hat{G}(\vk,i\omega_n)&=&0. \nonumber \\
\end{eqnarray}
The two-band  structure of the normal state is more conveniently
handled, if we use the Green's functions $G_{1,2}$ and $F_{1,2}$,
as is easily obtained by the unitary transformation
(\ref{Utrans}),
\begin{eqnarray}
    \left( \begin{array}{cc}
    G_{1} & 0\\
    0 & G_{2} \\
    \end{array} \right) =\hat{U}^{\dag}(\vk)\,\hat{G} \,\hat{U}(\vk)\nonumber \\
    -\left( \begin{array}{cc}
    F_{1} \, e^{-i\phi_{\vsk}} & 0\\
    0 & F_{2} \, e^{i\phi_{\vsk}} \\
    \end{array} \right) =\hat{U}^{\dag}(\vk)\,\hat{F} \,\hat{U}^{*}(-\vk).\nonumber \\
\end{eqnarray}
The phase factor $-e^{\mp i\phi_{\vsk}}$ is introduced to cancel
the phase dependence of the anomalous Green's functions on the two
bands \cite{Sergienko2004}.

The pairing interaction defined in (\ref{Inter}) fixes the form of
the order parameter to $\hat{\Delta}(\vk)=i \{\psi+d \,(\vg_{\vsk}
\cdot \hat{\vsig}) \} \hat{\sigma}_{y}$. $\psi$ and $d$ are
complex order parameters which can be interpreted as the magnitude
of the s-wave spin-singlet and respectively protected spin-triplet
components. The particular form of $\hat{\Delta}(\vk)$ prevents
the existence of inter-band terms in the Gor'kov equations
\begin{eqnarray}
    \label{Gorkov2}
    \left(\{G^{0}_{1,2}(\vk,i\omega_n)\}^{-1}-\Sigma_{G}(i\omega_{n}) \right) G_{1,2}(\vk,i\omega_n) \nonumber \\
    \quad \quad +\left(\Delta_{1,2}(\vk)+\Sigma_{F}(i\omega_{n})\right)  F^{{\dag}}_{1,2}(\vk,i\omega_n)&=&1 \nonumber \\
    \left(\{G^{0}_{1,2}(-\vk,-i\omega_n)\}^{-1}+\Sigma_{G}(-i\omega_{n})\right) F^{\dag}_{1,2}(\vk,i\omega_n) \nonumber \\
    \quad \quad
    -\left(\Delta_{1,2}^{*}(\vk)+\Sigma_{F}(i\omega_{n})\right) G_{1,2}(\vk,i\omega_n)&=&0, \nonumber \\
\end{eqnarray}
where in this case
\begin{eqnarray}
    \label{Self}
    \Sigma_{G}(i \omega_{n})
    &=&\frac{\Gamma}{2 \pi N_{0}} \int  \frac{d\vk'}{(2\pi)^{3}} \left \{ G_{1}(\vk',i \omega_{n}) + G_{2}(\vk',i \omega_{n}) \right \}
    \nonumber \\
    \Sigma_{F}(i \omega_{n})&=&\frac{\Gamma}{2 \pi N_{0}} \int \frac{d\vk'}{(2\pi)^{3}} \left \{ F_{1}(\vk', i \omega_{n}) +  F_{2}(\vk', i \omega_{n}) \right \}, \nonumber \\
\end{eqnarray}
and,
\begin{equation}
\label{GapEq} \Delta_{1,2}(\vk)=(\psi \pm d |\vg_{\vk}|).
\end{equation}
Thus the Gor'kov equations are diagonal in the band index.
Interband effects are contained in the gap function and the
scattering self-energies. The scattering off an impurity does not
change the spin of a quasiparticle and with the impurity-average a
certain translational symmetry is restored, such that the two
bands to not mix in this approximation. Interband effects occur
only through virtual processes.


Introducing the modified gap functions $ \tDelta_{1,2}(\vk,i
\omega_{n})=\Delta_{1,2}(\vk)+\Sigma_{F}(i\omega_{n})$ and
frequencies $ i  \tomega_n=i \omega_n-\Sigma_{G}(i\omega_{n})$,
the solution of the two-band Gor'kov equations is given by
\begin{equation}
\label{GreenG} G_{1,2} ({\bm k}, i \omega_n) =- \frac{i
\tomega_{n}+\xi_{1,2}
}{(\tomega_{n}^{2}+|\tDelta_{1,2}|^{2}+\xi^{2}_{1,2})},
 \end{equation}
and,
\begin{equation}
\label{GreenF} F_{1,2}  ({\bm k}, i \omega_n) =
\frac{\tDelta_{1,2}
}{(\tomega_{n}^{2}+|\tDelta_{1,2}|^{2}+\xi^{2}_{1,2})}.
\end{equation}
and $ \tDelta_{1,2}(\vk,i \omega_{n})$ and $ \tomega_n $ have to
be determined self-consistently. Going back to the spin basis we
find
\begin{eqnarray}
\label{Green}
\hat{G} (\vk, i \omega_n) &=& G_{1} (\vk, i \omega_n) \hat{\sigma}_{1}(\vk)+G_{2} (\vk, i \omega_n) \hat{\sigma}_{2}(\vk), \nonumber \\
\hat{F} (\vk, i \omega_n) &=& \{ F_{1} (\vk, i \omega_n) \hat{\sigma}_{1}(\vk)+F_{2} (\vk, i \omega_n) \hat{\sigma}_{2}(\vk) \} i \hat{\sigma}_{y}. \nonumber \\
\end{eqnarray}

\section{Characterization of the superconducting instability}

The gap functions of the two bands are obtained solving the
self-consistent equation
\begin{equation}
\label{GapEqn1} \Delta_{\alpha}(\vk)=- k_{B} T \int
\frac{d\vk'}{(2\pi)^{3}}  \sum_{n,\beta} V^{\alpha,
\beta}_{\vk,\vk'} F_{\beta}(\vk',i\omega_n).
\end{equation}
The corresponding pairing interaction is determined by
\begin{eqnarray}
 V^{\alpha, \beta}_{\vk,\vk'}&=& \sum_{\gamma,\delta} \sum_{s_{i},s'_{i}} D^{*}_{\alpha,\gamma}(\vsk) U^{*}_{s_{1} \gamma} (-\vsk) \hat{U}^{\dag}_{\gamma s_{2}}(\vsk) V_{s_{1}s_{2},s'_{2}s'_{1}}(\vsk, \vsk') \nonumber \\
&\times&\hat{U}_{s'_{2}\delta}(\vsk') \hat{U}^{\top}_{\delta
s'_{1}}(-\vsk') D_{\delta, \beta}(\vsk'),
\end{eqnarray}
with
$\hat{D}(\vsk')=$diag$(-\exp(-i\phi_{\vsk'}),-\exp(i\phi_{\vsk'}))$.
Using (\ref{Inter}) and keeping only the contributions to pairing
in the $A_1$ pairing channel, it follows
\begin{eqnarray}
       \label{InterSpinor}
    \hat{V}_{\vk,\vk'}&=& \frac{V}{2} \left\{ \left[ e_{s}+e_{t}|\vg_{\vsk}||\vg_{\vsk'}| \right] \hat{\sigma}_{0}
+ \left[e_{s}-e_{t} |\vg_{\vsk}||\vg_{\vsk'}| \right] \hat{\sigma}_{x} \right. \nonumber \\
    &-& \left. e_{m} \left[|\vg_{\vsk}|+|\vg_{\vsk'}| \right] \hat{\sigma}_{z}
    - i e_{m} \left[|\vg_{\vsk}|-|\vg_{\vsk'}| \right] \hat{\sigma}_{y}
    \right\}. \nonumber \\
\end{eqnarray}
We parameterize the distribution of the normal state density of
states in the two bands by constants $ N_1 = N_0 (1 + \delta_N) $
and $ N_2 = N_0 (1 -  \delta_N) $. To lowest order approximation
one may consider $ \delta_N \approx \alpha N_0' (\xi =0) / N_0$.

%

The critical temperature and the basic structure of the gap
function $ \vDelta = (\psi, d) $
 which characterizes the superconducting instability, follow from the solution of the linearized form of the
self-consistent system of Eqs.
(\ref{SelfConsist1}-\ref{SelfConsist3}). Introducing Eqn.
(\ref{SelfConsist1}) in Eqn. (\ref{SelfConsist3}), and using
standard summation techniques, we find
\begin{equation}
\label{SelfConsistLin}
    \frac{1}{N_{0}V} \vDelta = \left\{f_{1}(\epsilon_{c},k_{B}T) \hat{Q}^{l}_{0} +f_{2}(\Gamma,k_{B}T) \hat{Q}^{l}_{\Gamma} \right \}\vDelta ,
\end{equation}
with
\begin{eqnarray}
f_{1}(\epsilon_{c},k_{B} T)&=& \ln(\epsilon_{c}/2 \pi k_{B}T)+\ln(4 \gamma) \nonumber \\
f_{2}(\Gamma,k_{B} T)&=& \Psi \left(\frac{1}{2}+\frac{\Gamma}{2 \pi k_{B} T}\right)-\Psi \left(\frac{1}{2}\right) \nonumber \\
\end{eqnarray}
and,
\begin{eqnarray}
\hat{Q}^{l}_{0}&=&
     \left[ \left( \begin{array}{cc} -e_{s} & e_{m}  \\
       e_{m} & -e_{t}
    \end{array} \right) +  \delta_{N} \langle |\vg_{\vsk}| \rangle \left( \begin{array}{cc} e_{m} & -e_{s}  \\
       -e_{t} & e_{m}
    \end{array} \right) \right], \nonumber \\
\hat{Q}^{l}_{\Gamma}&=& (1-\delta^{2}_{N} \langle |\vg_{\vsk}| \rangle^{2}) \left( \begin{array}{cc} 0 &  -e_{m}   \\
       0 & e_{t}
    \end{array} \right).
\end{eqnarray}
Here we have introduced the digamma function $\Psi(z)$ defined by
$ \Psi(z) \equiv d /dz \ln(z!)$, and the Euler's constant $\ln
\gamma=C \approx 0.577$. For simplicity we have assumed the same
cut-off energy $\epsilon_{c}$ for both bands.

First we characterize the superconducting instability of the clean
system, i.e. $ \Gamma =0 $. Eqn. (\ref{SelfConsistLin}) has a
non-trivial solution, if at least one of the eigenvalues
$\lambda_{i}$ of the matrix $\hat{Q}^{l}_{0}$ is positive.  In
this case the critical temperature follows the standard BCS
relation $k_{B}T_{c}=2 \epsilon_{c} \gamma/\pi \; \exp(-1/(N_{0}V
\lambda'))$, where $\lambda'=\max_{i}(\lambda_{i}) > 0$.
Furthermore the nucleating form of the gap function follows from
$\hat{Q}^{l}_{0} \vDelta'=\lambda' \vDelta'$.

In this context we introduce the term of {\it dominant} channel to
denote the channel responsible for the superconducting transition,
and we  call {\it subdominant} the other channel characterized by
$\lambda''=\min_{i}(\lambda_{i})$ and $\hat{Q}^{l}_{0}
\vDelta''=\lambda'' \vDelta''$.

Solving the eigenstate problem $\hat{Q}^{l}_{0}
\vDelta_{i}=\lambda_{i} \vDelta_{i}$, we find  the sets
\begin{eqnarray}
\label{lambda1}
2 \lambda'&=&  -(e_{s}+e_{t})+2 e_{m} \delta + \Lambda   \\
\label{Delta1}
\vDelta' &\propto& \left( \begin{array}{c} -(e_s-e_t)+\Lambda \\
2(e_{m}-e_{t} \delta) \end{array} \right) ,
\end{eqnarray}
and
\begin{eqnarray}
\label{lambda2}
2 \lambda''&=& -(e_{s}+e_{t})+2 e_{m} \delta - \Lambda,  \\
\label{Delta2}
\vDelta'' &\propto& \left( \begin{array}{c} -(e_s-e_t)-\Lambda \\
2(e_{m}-e_{t} \delta) \end{array} \right).
\end{eqnarray}
where
\begin{eqnarray}
\Lambda=
\sqrt{(e_{s}-e_{t})^{2}+4(e_{m}-e_{s}\delta)(e_{m}-e_{t}\delta)},
\nonumber
\end{eqnarray}
and $\delta= \delta_{N} \langle |\vg_{\vsk}| \rangle$ to simplify
the notation. Our interest lies on the characterization of the
instability for all possible combinations of the three components
of the pairing potential (\ref{Inter}).   From Eqn.
(\ref{lambda1}) we determine the conditions for the
superconducting instability and the form of the nucleating pairing
state given by Eq. (\ref{Delta1}).

\begin{figure}[h]
\begin{center}
 \includegraphics[width=8.5cm, height=4.5cm ]{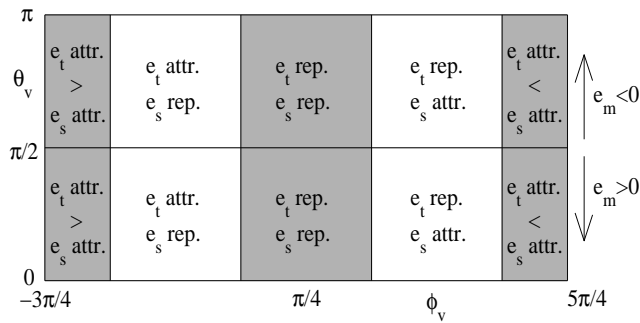}
\begin{minipage}[c]{8cm}
\caption{\label{InterFig0} The character of the three interaction
components ($e_{s} \equiv$ s-wave spin-singlet, $e_{t} \equiv$
protected spin-triplet, and $e_{m} \equiv$ parity violating) as a
function of the two spherical angles ($\theta_{v}, \phi_{v}$). The
abbreviations '' $e_{s}$ attr. '' and '' $e_{s}$ rep.'' mean that
the s-wave component of the interaction is attractive ($e_{s} <
0$) or respectively repulsive  ($e_{s} > 0$).}
\end{minipage}
\end{center}
\end{figure}

In order to conveniently display a phase diagram of the three
parameters we use the condition $ e_s^2 +e_t^2 + e_m^2 =1 $ and
represent these parameters by the angles of spherical coordinates
($e_{s} \equiv$ s-wave spin-singlet, $e_{t} \equiv$ $S$
spin-triplet, and $e_{m} \equiv$ mixing):
\begin{eqnarray}
\label{param}
e_{s}&=&\cos(\phi_{v})\sin(\theta_{v}), \nonumber \\
e_{t}&=&\sin(\phi_{v})\sin(\theta_{v}), \nonumber \\
e_{m}&=&\cos(\theta_{v}).
\end{eqnarray}
The character of the interaction as a function of the two
spherical angles ($\theta_{v}, \phi_{v}$) is shown in Fig.
\ref{InterFig0}.

\begin{figure}[h]
\begin{center}
 \includegraphics[width=8.5cm, height=4.5cm ]{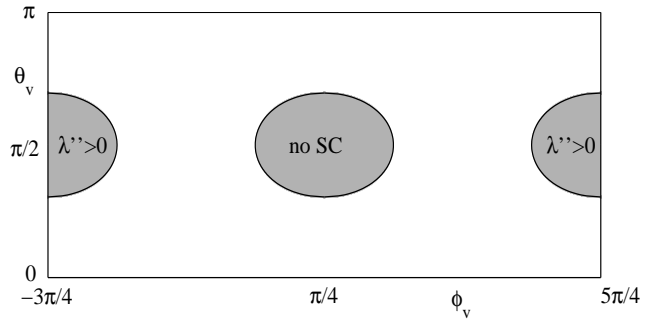}
\begin{minipage}[c]{8cm}
\caption{\label{InterFig1} The character of the eigenvalues
$\lambda'$, Eqn. ( \ref{lambda1}),  and $\lambda''$ , Eqn. (
\ref{lambda2}), as a function of the two spherical angles
($\theta_{v}, \phi_{v}$). If $\lambda' < 0$ we don't have any
superconducting instability (grey domain with ''no SC''). The
white domain is characterized by $\lambda' > 0$, and $\lambda'' <
0$. If $\lambda'' > 0$, the system is characterized by two
critical temperature (grey domain with '' $\lambda'' > 0$'') }
\end{minipage}
\end{center}
\end{figure}

Obviously an attractive interaction either for spin-singlet or
-triplet pairing or both would yield a superconducting
instability. If both $ e_s $ and $ e_t $ are repulsive ($>0$), the
mixing term plays the decisive role. From Eq. (\ref{lambda1}) we
see immediately that an instability is absent for small $ | e_m |
$ (grey domain with ''no SC'' in Fig. \ref{InterFig1}) :
\begin{equation}
\label{NSC} |e_{m}| < \sqrt{e_{s}e_{t}} \textrm{ where } e_{s} ,
\;e_{t} > 0.
\end{equation}
On the other hand, a large enough value of $ | e_m | $ can trigger
superconductivity, even with $ e_s , e_t > 0$. The inter-parity
scattering can lower the pairing energy and generate positive
eigenvalue of the matrix $\hat{Q}^{l}_{0}$. This mechanism is
formally analog to superconductivity driven by interband Cooper
pair scattering in a multi-band superconductor.

A second positive eigenvalue, $ \lambda '' > 0 $ is possible, if
both spin channels are attractive ($ e_s, e_t < 0 $) (grey domain
with ''$e_{s}$ and $e_{t}$ attr.'' in Fig. \ref{InterFig0}).
From Eqn. (\ref{lambda2}) we see that the following condition has to be satisfied:
\begin{equation}
|e_{m}| < \sqrt{e_{s}e_{t}} \textrm{ and } e_{s} , \;e_{t} < 0.
\end{equation}
The parity-mixing term tends to suppress the second instability.

It remains only to determine the form of the 2-dimensional order
parameter $\Delta=(\psi,d)$ based on Eqn. (\ref{Delta1}), which
nucleates at the superconducting transition. There are four
typical forms of the order parameter $ ( \psi, d )$, which will
help us to structure the following discussion ($ \vDelta = (\psi,
d) $):
\begin{eqnarray}
\label{DirDelta}
\vDelta_{s}&\propto&  \left( \begin{array}{c} 1    \\
       0  \end{array} \right) \textrm{Spin-singlet}, \quad
\vDelta_{t} \propto  \left( \begin{array}{c} 0    \\
       1  \end{array} \right) \textrm{Spin-triplet}, \nonumber \\
\vDelta_{m1}&\propto&  \left( \begin{array}{c} 1    \\
       1  \end{array} \right) \textrm{ 1st band },  \quad
\vDelta_{m2}\propto  \left( \begin{array}{c} 1    \\
       -1  \end{array} \right) \textrm{2nd band}.
\end{eqnarray}
These correspond to the pure spin-singlet pairing $\vDelta_{s}$,
the pure spin-triplet pairing $\vDelta_{t}$ and two extreme cases
of the mixed pairings $\vDelta_{m1}$, and $\vDelta_{m2}$ .

\begin{figure}[h]
\begin{center}
 \includegraphics[width=8cm, height=8cm ]{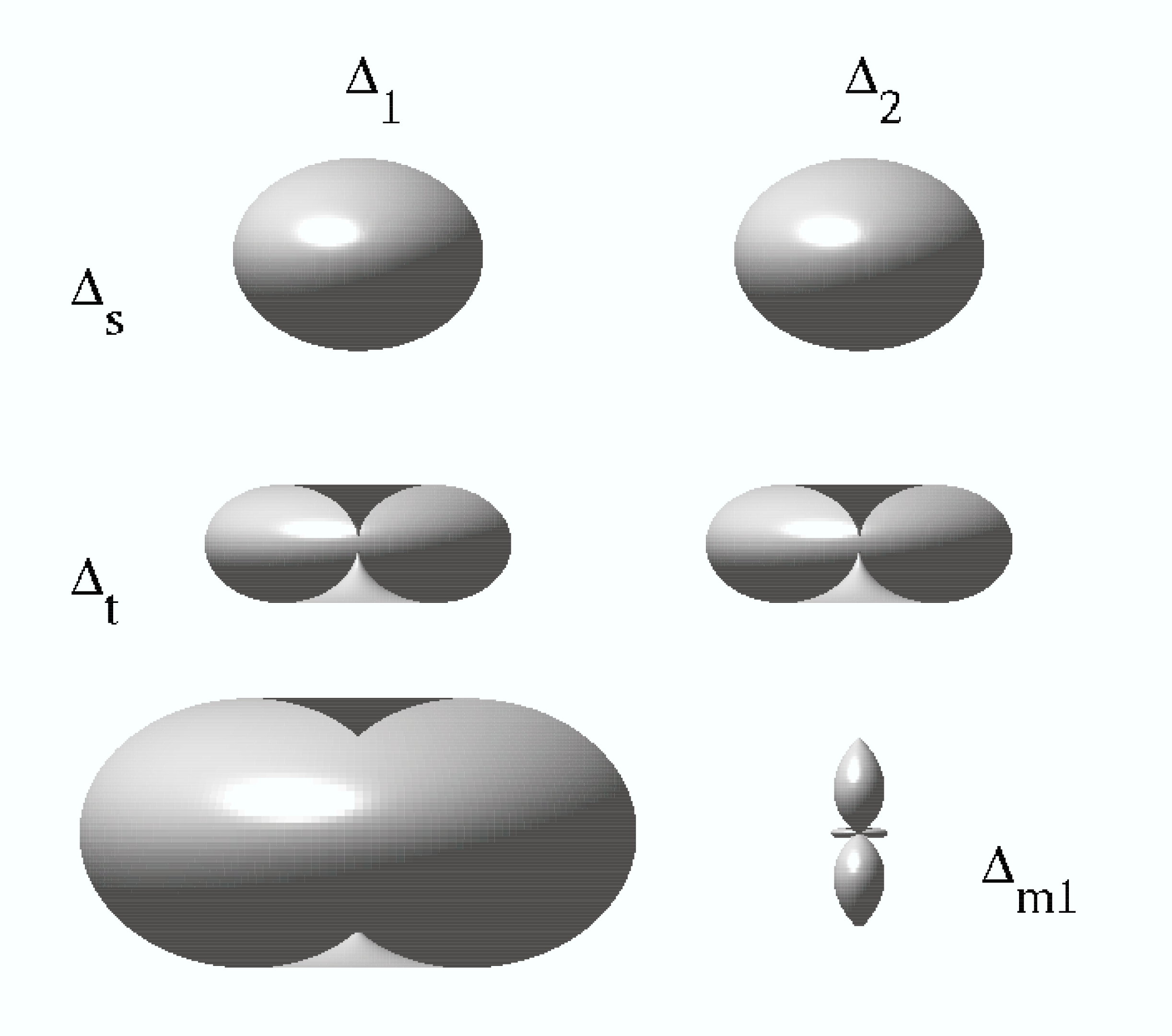}
\begin{minipage}[c]{8cm}
\caption{\label{Gaps} The gap amplitudes $\Delta_{1,2}$ of the two
non degenerated bands for three particular directions of the order
parameter in the 2-dimensional plane, spanned by $\psi$, $d$. The
directions corresponding to the two dimensional order parameter
$\vDelta_{s}$, $\vDelta_{t}$, $\vDelta_{m1}$ are given by
(\ref{DirDelta}) }
\end{minipage}
\end{center}
\end{figure}

Before we start to discuss the conditions favoring different order
parameter forms, we consider the corresponding gap topologies on
the two bands for the four cases in Eq. (\ref{DirDelta}). To be
concrete we discuss now the case relevant for the CePt$_{3}$Si,
ie. $\vg_{\vsk} \propto (k_{y}, -k_{x}, 0)$. The gap on the two
Fermi surfaces is given by
\begin{equation}
\label{Gaps1} \Delta_{1,2}(\theta)=(\psi \pm d
\sqrt{3/2}|\sin(\theta)|),
\end{equation}
where $\theta$ is the polar angle in $\vk$-space relative the
$z$-axis. Fig. \ref{Gaps} displays the two bands gap function
$\Delta_{1,2}$ for the three different order parameter
$\vDelta_{s}$, $\vDelta_{t}$, $\vDelta_{m1}$ of Eq.
(\ref{DirDelta}). The case $\vDelta_{m2}$ can be obtained from
$\vDelta_{m1}$  by exchanging  $\Delta_{1}$ and $\Delta_{2}$.

In the cases of the pure spin-singlet phase $\vDelta_{s}$ or the
pure spin-triplet phase $\vDelta_{t}$, the two gap functions
$\Delta_{1}$ and $\Delta_{2}$ have the same amplitude, with the
difference that in the spin-triplet case point-nodes appear along
the z-axis. For $\vDelta_{m1}$ the gap on the first Fermi surface
is much bigger  than on the  second, $\Delta_{1} > \Delta_{2}$.
Furthermore the gap function in the second band $\Delta_{2}$ has
two line-nodes perpendicular to the z-axis. The presence of
line-nodes in one of the two non-degenerate bands is not a
particularity of the mixed states $ \vDelta_{m1} $, but a general
property of all states with $|d| C_{1} >|\psi|$ and $|\psi|
\neq0$, where $C_{1}=\sqrt{3/2}$ only in case of a spherical
original Fermi surface. However in general the topology of the
nodes in the two gap functions belonging to the trivial
representation $A_{1}$ of the point group $C_{4v}$ is
characterized by
\begin{eqnarray}
\label{Top}
\frac{|\psi|}{|d|} \left\{\begin{array}{cc} =0 & \textrm{2 point-nodes} \\
< C_{1} & \textrm{2 line-nodes for one band} \perp \textrm{z-axis} \\
> C_{1} & \textrm{no nodes}
\end{array} \right.
\end{eqnarray}
%

Now we turn to the conditions under which the different forms of
the order parameter nucleate. We expect that the order parameter
appears as pure spin-singlet $\Delta_{s}$ or a pure spin-triplet
$\Delta_{t}$, if the spin-singlet potential $e_{s}$ or the
spin-triplet potential $e_{t}$ are attractive and dominant,
respectively. This is the case if the density of states on the
Fermi levels is equal $\delta_N=0$ and the parity-mixing
interaction $e_{m}$ is absent.

From Eq. (\ref{Delta1}) the necessary conditions can be derived
\begin{eqnarray}
\label{Cond1} \label{singletline}
\vDelta=&\vDelta_{s} &  \mbox{for} \quad e_{m}=e_{t} \delta , \;  e_{s} < e_{t},  \\
\label{tripletline}
\vDelta=&\vDelta_{t}  &   \mbox{for} \quad e_{m}=e_{s}  \delta,  \;  e_{t} < e_{s},  \\
\label{mix1line}
\vDelta=&\vDelta_{m1} &   \mbox{for} \quad e_{s} =  e_{t} , \;  e_{m} > e_{t} \delta,  \\
\label{mix2line} \vDelta=&\vDelta_{m2} &  \mbox{for} \quad e_{s} =
e_{t} , \;  e_{m} < e_{t}. \delta
\end{eqnarray}
Moreover the condition in Eq. (\ref{NSC}) has to be satisfied to
guarantee a superconducting phase transition. The first two
conditions (\ref{singletline}) and (\ref{tripletline}) are
necessary to compensate the  mixing of the spin-triplet with the
spin-singlet pairing channels induced when the DOS are different
for the two Fermi surfaces $\delta \neq 0$. Furthermore taking
into account of the condition (\ref{NSC}) we see that
 surprisingly
 $\vDelta_{s}$ and $\vDelta_{t}$ can also nucleate in presence of repulsive spin-singlet $e_{s}$ and spin-triplet $e_{t}$ potentials.

If $e_{s}=e_{t}$ the conditions (\ref{mix1line}) and
(\ref{mix2line}) indicate that the order parameter is either
$\vDelta_{m1}$ or $\vDelta_{m2}$. For $\delta_N=0$ the sign of the
parity-mixing potential  $e_{m}$  determines in which of the two
order parameter appears.

\begin{figure}[h]
\begin{center}
 \includegraphics[width=8.5cm, height=6cm ]{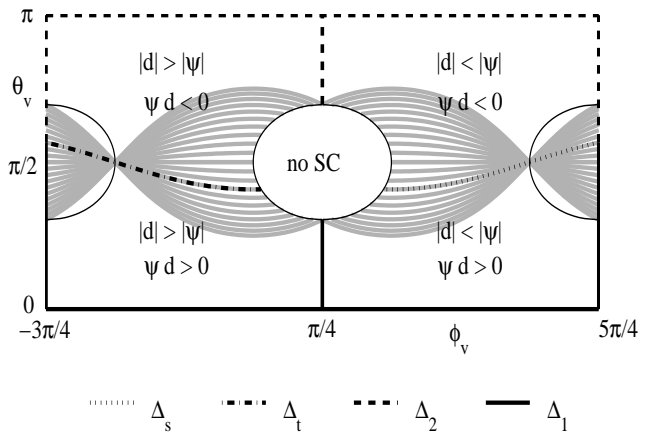}
\begin{minipage}[c]{8cm}
\caption{\label{InterFig2} \small{The direction of the nucleated
order parameter $\vDelta=(\psi,d)$. The grey lines show different
possible lines for $\vDelta=\vDelta_{s}$, and,
$\vDelta=\vDelta_{t}$. The  black ones are obtained for
$\delta=0.3$.}}
\end{minipage}
\end{center}
\end{figure}
Fig. \ref{InterFig2} shows the phase diagram of the different
order parameter forms resulting from Eq. (\ref{DirDelta}). The
different black lines are plotted choosing $\delta=0.3$. The  grey
ones show how the lines corresponding to the pure spin-singlet
$\Delta_{s}$ and pure spin-triplet $\Delta_{t}$ are moved varying
$\delta$. For general combinations of the three components
$e_{s}$, $e_{t}$, $e_{m}$ the angles $\theta_{v}, \phi_{v}$ lie in
one of the four basic domains delimited by the black lines shown
in Fig. \ref{InterFig2}. The order parameter nucleates with
intermediary values of $\psi$, and $d$.

\section{The effect of the disorder}

We now turn to the superconducting instability in a disordered
system and analyze the behavior of the critical temperature and
the structure of the order parameter for different coupling
parameters and impurity concentrations. For this purpose we
examine now Eq. (\ref{SelfConsistLin}) for $\Gamma >0$. First we
would like to note that the zeros in the first column of the
matrix  $\hat{Q}^{l}_{\Gamma}$ are a consequence of Anderson's
theorem \cite{Anderson1959}, i.e. the conventional $s$-wave
pairing state is not affected by non-magnetic impurities. For more
general states in our two-dimensional order parameter space $ T_c$
decreases with growing disorder. We distinguish two basically
different cases here. For the pure system either both eigenvalues
$\lambda', \lambda''$ are positive, or one of the two, $ \lambda''
$ is negative. In the first case there is a second lower (bare)
critical temperature $k_{B}T''_{c}=1.14 \epsilon_{c} \;
\exp(-1/(N_{0}V \lambda''))$, indicating the possibility of second
superconducting phase transition.

\subsection{Case: $ \lambda''=0 $}

Before starting the discussion of this two general cases we
consider the boundary situation with $\lambda''=0$.  The
interaction parameters $e_{s}$, $e_{t}$, and, $e_{m}$ lie on a
line given by,
\begin{equation}
\label{circle} |e_{m}| = \sqrt{e_{s}e_{t}} \textrm{ with } e_{s} ,
\;e_{t} < 0,
\end{equation}
which correspond to the circle centered in $\phi_{v}=5 \pi /4
\equiv -3 \pi/4$ in the phase diagram of Figs. \ref{InterFig1},
and, \ref{InterFig2}. The only eigenvector of the matrix
$\hat{Q}^{l}_{\Gamma}$, with a non-vanishing eigenvalue
corresponds to the form of $\vDelta'$ given by in Eq.
(\ref{Delta1}).  Hence in this case disorder would not alter the
structure of the nucleating order parameter
\begin{eqnarray}
\label{DeltaD1}
\vDelta &=&\left( \begin{array}{c} \psi  \\
 d \end{array} \right) \propto \vDelta' \propto \left( \begin{array}{c} \sqrt{|e_{s}|}  \\
 \textrm{sign}(e_{m}) \sqrt{|e_{t}|} \end{array} \right).
\end{eqnarray}
The instability equation (\ref{SelfConsistLin}) becomes
\begin{equation}
\label{SelfConsistLin1}
    \frac{1}{N_{0}V} = \lambda' f_{1}(\epsilon_{c}, k_{B} T)  + e_{t}(1-\delta^{2})f_{2}(\Gamma, k_{B} T),
\end{equation}
where $e_{t}(1-\delta^{2})$ is the eigenvalue of the matrix
$\hat{Q}^{l}_{\Gamma}$. We replace
$f_{1}=\ln(T_{c}/T)+1/N_{0}V\lambda'$ and
$\lambda'=-(e_s+e_t)+2e_{m}\delta $ in Eq. (\ref{SelfConsistLin1})
\begin{equation}
\label{SelfConsistLin2}
    \ln(T_{c}/T)= \frac{e_{t}(1-\delta^{2})}{(e_s+e_t)-2e_{m}\delta}f_{2}(\Gamma, k_{B} T),
\end{equation}
Using Eq. (\ref{DeltaD1}) we obtain
\begin{equation}
\label{TcPure}
    \ln\left(\frac{1}{t}\right)=\frac{1}{1+\eta} f_{2}(\gamma,t),
\end{equation}
where
\begin{equation}
\label{eta}
    \eta=\frac{(\psi+\delta d)^{2}}{{d}^{2}(1-\delta^{2})}.
\end{equation}
with $t=T/T_{c}$ and $\gamma=\Gamma/k_{B}T_{c}$. The effect of
impurity scattering affects $ T_c $ as in unconventional
superconductors in general \cite{Larkin1965}. However, a
distinctive point is the presence of the pre-factor $1/(1+\eta)$.
A similar result was obtained for the s+g-superconductivity in
borocarbides \cite{Yuan2003}.

In order to visualize the behavior of the onset temperature of
superconductivity depending on impurity concentrations, we
introduce the normalization with respect to the initial slope of $
T_c$-reduction.
\begin{equation}
    n'=\frac{n_{imp}}{(-d n_{imp}/d t)_{t=1}}=\frac{\tilde{\gamma}}{(-d \tilde{\gamma}/d t)_{t=1}}.
\end{equation}

\begin{figure}[h]
\begin{center}
 \includegraphics[width=7.5cm, height=6cm ]{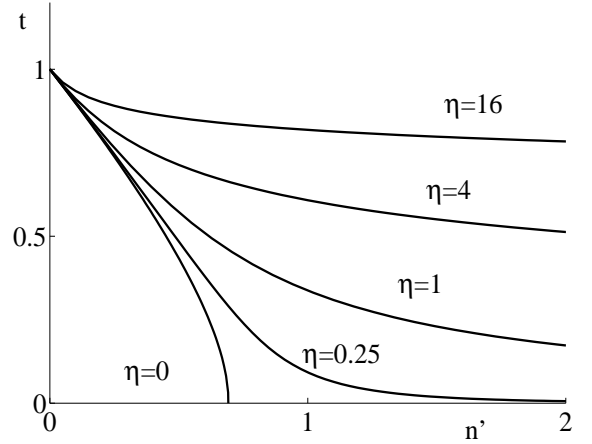}
\begin{minipage}[c]{8cm}
\caption{\label{TcEtaFig}  The critical temperature as a function
of the impurity concentration reduced so as to have a slope of
$-1$ at $t=1$ for different value of $\eta$. The value of $\eta$,
see eqn. (\ref{eta}). This result is valid only for
$\lambda''=0$.}
\end{minipage}
\end{center}
\end{figure}

Fig. \ref{TcEtaFig} shows the evolution of $t$ as a function of
$n'$ for different values of $\eta$ when the subdominant channel
has $ \lambda''=0$.

For $\eta > 0$ the superconducting instability extends to large
impurity concentrations $n'$ with the asymptotical behavior of the
critical temperature
\begin{equation}
    0.88 \, t^{-\eta}-t= \frac{4(1+\eta)}{\pi} n'.
\end{equation}
Hence we observe a variation of the robustness against
non-magnetic impurities. For $ \eta = 0 $ the standard behavior of
an unconventional superconductor is found, while for $ \eta \to
\infty $ the non-sensitivity against disorder analogous to a
conventional superconducting phase is realized. The latter case
coincides with the suppression of the spin-triplet component in
the pure phase.
The exponent $\eta$ can be interpreted as the product of two
ratios
\begin{equation}
    \eta=\eta_{1}\eta_{2} \textrm{ with } \eta_{1,2}=\frac{\langle N_{1}\Delta_{1}(\vsk)+N_{2}\Delta_{2}(\vsk)\rangle}{N_{1,2}\langle\Delta_{1}(\vsk)-\Delta_{2}(\vsk)\rangle}.
\end{equation}
where $\eta_{1,2}$ correspond to the ratio between the s-wave
spin-singlet component (numerator) and the spin-triplet component
lying in the first and respectively second band (denominator).
Once $\langle N_{1}\Delta_{1}(\vsk)+N_{2}\Delta_{2}(\vsk)\rangle
\neq 0$ the robustness of the superconducting state against
disorder is introduced by the presence of a s-wave spin-singlet
contribution.
For other unconventional pairing states which do not belong to the trivial representation $A_{1}$, the impurities contribution to the self energy of anomalous type is zero and we recover exactly the result obtained by Larkin \cite{Larkin1965}. 

 \subsection{Case: $ \lambda'' \neq 0 $}

In this case the eigenvectors of the two matrices
$\hat{Q}^{l}_{0}$ and $\hat{Q}^{l}_{\Gamma}$ are different, such
that we need to diagonalize the matrix appearing in Eq.
(\ref{SelfConsistLin}). The highest eigenvalue and its eigenvector
read
\begin{eqnarray}
\label{lambda1g}
2 \lambda &=& 2 \lambda' \, f_{1} +c_{1}-c_{2}+\sqrt{c^{2}_{2}+c_{3}}, \\
\label{Delta1g}
\vDelta & \propto & \left( \begin{array}{c} \psi'+\frac{\sqrt{c^{2}_{2}+c_{3}}-c_{4}}{f_{1} c_{5}} \\
d' \end{array} \right),
\end{eqnarray}
where $(\psi',d')$ are normalized ($\psi'^{2}+d'^{2}=1$).
Furthermore, the parameters $ c_i $ are defined as
\begin{eqnarray}
c_{1}&=&(1-\delta^{2})\left[e_{t}-\frac{e_{t}(e_{t}-e_{s}-2 e_{m} \delta)+2 e_{m}^{2}}{\Lambda} \right]\, f_{2}, \nonumber \\
c_{2}&=&\Lambda \, f_{1} -\frac{(1-\delta^{2})}{2}\left\{\Lambda+\frac{e^{2}_{t}+e_{s}[4\delta(e_{m}-e_{t}\delta)-e_{s}]}{\Lambda} \right\}  f_{2}, \nonumber \\
c_{3}&=& -\frac{4(e^{2}_{m}-e_{s}e_{t})(e_{m}-e_{t}\delta)^{2}(\delta^{2}-1)^{2}}{\Lambda^{2}} f^{2}_{2}, \nonumber \\
c_{4}&=& \Lambda f_{1}+e_{t}(1-\delta^{2}) f_{2}, \nonumber \\
c_{5}&=& \sqrt{4(e_{m}-e_{t}\delta)^{2}+(e_{t}-e_{s}+\Lambda)^{2}}. \nonumber \\
\end{eqnarray}

The equation determining the critical temperature can be expressed
as
\begin{equation}
\label{EqnLambdaNZ1}
    \frac{2}{N_{0}V} = 2\lambda' f_{1}(\epsilon_{c}, k_{B} T)  +c_{1}-c_{2}+\sqrt{c^{2}_{2}+c_{3}}.
\end{equation}
When we use again the relation
$f_{1}=\ln(T_{c}/T)+1/N_{0}V\lambda'$, we can cancel the
$2/N_{0}V$ term appearing at the left hand side of Eq.
(\ref{EqnLambdaNZ1}) and reach a new convenient representation,
\begin{equation}
\label{EqnLambdaNZ2}
     \left[2 \lambda' \ln\left(\frac{1}{t}\right)+c_{1}\right]^{2} -2\left[2 \lambda' \ln\left(\frac{1}{t}\right)+c_{1}\right]c_{2}-c_{3}=0.
\end{equation}
We assume now that the second channel is attractive $\lambda''>0$,
so that it is possible to find a second solution for the
instability temperature $t''=T''_{c}/T_{c}$. We use the relation
$f_{1}=\ln(T_{c}/T)+\ln(t'')\lambda''/(\lambda''-\lambda')$ to
simplify Eqn. (\ref{EqnLambdaNZ2}).

This leads to an equation of second order in $\ln(t)$ of the
following form which allows us to determine both transition
temperatures ($t=T/T_{c}$ or $t=T''/T_{c}$)
\begin{equation}
\label{TcT2}
    \ln\left(\frac{1}{t}\right)\left( \ln(t'') -f_{2}(\gamma,t)-\ln(t) \right)= \frac{1}{1+\eta} f_{2}(\gamma, t)\ln(t''), \nonumber \\
\end{equation}
where now $\eta$ is given by the form of the order parameter in
the clean system
\begin{equation}
\label{eta'}
    \eta=\frac{(\psi'+\delta d')^{2}}{{d'}^{2}(1-\delta^{2})},
\end{equation}
and  $(\psi',d')$ are connected to the pairing potential via Eqn.
(\ref{Delta1}).

\begin{figure}[h]
\begin{center}
 \includegraphics[width=7.5cm, height=6cm ]{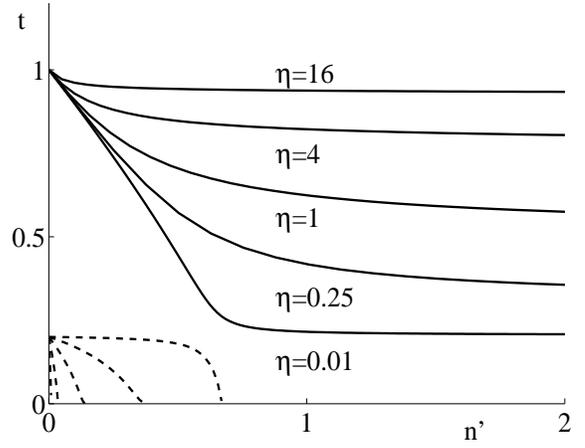}
\begin{minipage}[c]{8cm}
\caption{\label{TcEtaFig2} The critical temperature as a function
of the impurity concentration and for different values of $\eta$.
Those corresponding to the dominant channel, solid black lines,
have been reduced so as to have a slope of $-1$ at $t=1$. The
shadow black lines show the evolution of the subdominant channel.
This result is obtained supposing that the subdominant channel is
attractive and $t''=0.2$.}
\end{minipage}
\end{center}
\end{figure}

Fig. \ref{TcEtaFig2} shows the solutions of Eqn. (\ref{TcT2}) for
different values of $\eta$ and for $t''_0=0.2$ in the pure case.

The solid line shows the onset temperature of superconductivity as
a function of $n'$. The dashed line correspond to the bare
transition temperature of subdominant instability, as obtained
from the square root appearing in Eq. (\ref{EqnLambdaNZ1}).

The presence of an attractive subdominant channel supports the
survival of superconductivity under non-magnetic impurity
scattering.  The lowest limit for the critical temperature is
given by
\begin{equation}
\label{TcT2limit}
    \lim_{n' \rightarrow \infty} t=t_0''^{\frac{1}{1+\eta}}. \nonumber \\
\end{equation}
Using Eq. (\ref{EqnLambdaNZ1}) to simplify Eq.(\ref{Delta1g}) we
find that the form of the order parameter nucleating at the
critical temperature $t$ is given by
\begin{eqnarray}
\label{DeltaT2}
\left( \begin{array}{c} \psi \\
 d \end{array} \right)&\propto & \left( \begin{array}{cc} 1+f_{3}(t) & \delta f_{3}(t) \\
 0 & 1 \end{array} \right) \left( \begin{array}{c} \psi' \\
 d' \end{array} \right)
\end{eqnarray}
where we introduced the function $ f_3(t) $ defined as
\begin{equation}
\label{f3}
f_{3}(t)=-\frac{(1+\eta)\ln(t)}{(1+\eta)\ln(t)-\ln(t'')}.
\end{equation}
The spin-singlet component of the order parameter increases with
disorder. In the limit of dirty system the character of the order
parameter is purely spin-singlet. Note that the subdominant
instability is in any case suppressed by the disorder. The larger
$ \eta $ the stronger the suppression.

\begin{figure}[h]
\begin{center}
 \includegraphics[width=7.5cm, height=6cm ]{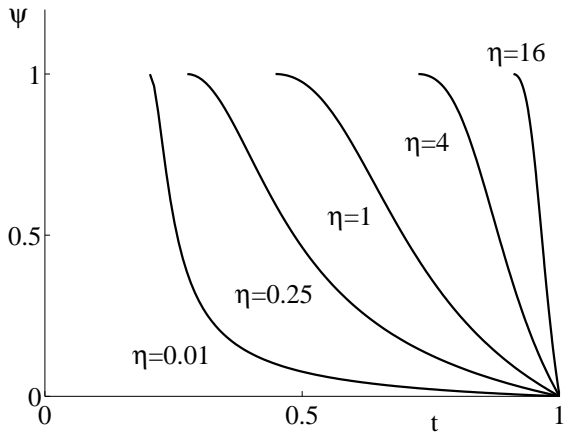}
\begin{minipage}[c]{8cm}
\caption{\label{PsiT2} The spin-singlet component $\psi$  of the
normalized order parameter as a function of the critical
temperature $t$ and for different values of $\eta$. This result is
obtained supposing that the subdominant channel is attractive with
$t''=0.2$ and that the order parameter of the clean system is
spin-triplet .}
\end{minipage}
\end{center}
\end{figure}

The behavior of the spin-singlet component $\psi$  of the
normalized order parameter as a function of the critical
temperature $t$ and for different values of $\eta$ is shown in the
Fig. \ref{PsiT2}. The order-parameter of the clean system is
supposed to be in the spin-triplet channel.

The case with a repulsive subdominant channel, $\lambda'' < 0$,
can be easily treated within the Eqs. (\ref{TcT2},\ref{f3})
substituting $\ln(t'')$ with $1/N_{0}V_{eff}$, where
\begin{equation}
\frac{1}{N_{0}V_{eff}}=\frac{1}{N_{0}V\lambda'}-\frac{1}{N_{0}V\lambda''}.
\end{equation}
From Eqn. (\ref{TcT2}) we find that superconductivity disappears for a sufficiently high impurity concentrations. The critical impurity concentration $n'_{c}$ is given by,
\begin{equation}
    0.88 \, e^{\frac{1}{N_{0}V_{eff}} \frac{\eta}{1+\eta}}= \frac{4(1+\eta)}{\pi} n'_{c}.
\end{equation}
Furthermore the spin-singlet component of the nucleating order
parameter decrease with disorder evolving towards
\begin{eqnarray}
\label{DeltaT2C}
\left( \begin{array}{c} \psi_{c} \\
 d_{c} \end{array} \right)&\propto & \left( \begin{array}{c} -\delta d' \\
  d' \end{array} \right)
\end{eqnarray}
at the critical impurity concentration $n'_{c}$.

It has to be noted that for almost  every combination of the
pairing interaction $\lambda'' \neq 0$. However the realization of
one of the two scenarios described in this chapter can be observed
only if $|\lambda''|$ is very close or respectively bigger than
$\lambda'$. This means than in general we will observe a behavior
very similar to that described by $\lambda''=0$.

\section{Conclusion}

In this paper we characterized the superconducting instability
towards a state belonging to the trivial representation $A_{1}$ in
a material without an inversion center and strong antisymmetric
spin-orbit coupling. The corresponding pairing state involves
spin-singlet s-wave pairing as well as a spin-triplet component
specific to the spin-orbit coupling. In addition to the pure
spin-singlet s-wave and spin-triplet pairing,  the pairing
interaction includes a parity mixing contribution corresponding to
an inter-parity scattering of Cooper pairs. The combination of the
three types of  pairing interaction ($e_{s}, e_{t}, e_{m}$), and
the distribution of the density of states on the two
non-degenerate bands $\delta$, determines the form of the order
parameter, represented by a singlet and a triplet component,
 $\Delta=(\psi,d)$. For the case of CePt$_{3}$Si, we found that if $|d| >  |\psi|$ and $|\psi| \neq 0$ ,
 then one Fermi surface would have accidental line nodes perpendicular to  the $z$-axis in the quasiparticle
 gap. The position of these accidental line nodes will in general
 depend upon both temperature and magnetic fields.

The two-component structure of the order parameter allows for two
distinct pairing channels, a dominant and a subdominant gap. We
have shown that depending on the properties characterizing the two
channels, the $A_1 $-phase is affected by non-magnetic impurities
in different ways. In all cases the dependence of $ T_c$ on the
impurity concentration is very characteristic and could be used to
establish the realization of the $A_1$-phase and the property  of
the subdominant pairing state. We would also like to mention that
the presence of line nodes may leave its specific low-temperature
features in various thermodynamic quantities such as the specific
heat, the London penetration depth and so on. Recent experimental
data of the nuclear magnetic relaxation rate $ T_1^{-1} $
\cite{Hayashi2005,Samokhin2005}
 and the London penetration depth give evidence for the possible presence of line nodes.
 Moreover, the presence of two instabilities in the linearized gap equation, dominant and subdominant,
 provide the possibility that two superconducting phase transitions could appear. From our study we
 conclude that such additional transitions would be observed in rather pure samples only, since the
 subdominant instability is very sensitive to disorder. This is compatible with the fact that in the dirty
 system eventually only the conventional pairing component would survive, while all alternative
 pairing channels are suppressed.

%
%
\section{Acknowledgments}
 The authors thank  A. Koga, and, T. M. Rice for useful discussions.
 This work was supported by the Swiss National Science Foundation and the NCCR
 MaNEP. DFA was also supported the National Science Foundation Grant No.
 DMR-0318665.

\appendix

\section{ \label{em} Origin of mixed singlet-triplet pairing interaction $e_m$}

In this Appendix we consider a microscopic origin for the mixed
singlet-triplet pairing interaction $e_m$ given in Eq.~13. In
particular, we show that that the Dzyaloshinskii-Moriya
\cite{Dzyaloshinskii1958, Moriya1960} (DM) magnetic interaction
gives a contribution to $e_m$. Such an interaction is well known
to exist for magnetic systems that break inversion symmetry. We
then show how this interaction can arise from the single particle
Hamiltonian given in Eq. 2 for both a weakly interacting Fermi
liquid and for a Hubbard model near half filling.  Note that the
DM interaction is not the only microscopic interaction that
contributes to $e_m$. Nevertheless, this discussion suffices to
show to $e_m$ is not zero in general and in some cases may provide
a substantial contribution  to the superconducting condensation
energy.

\subsection{DM Interaction}

The DM interaction can be written
\begin{equation}
H_{DM}=\frac{1}{N}\sum_{\vq}i\vD(\vq)\cdot
\vS_{\vq}\times\vS_{-\vq} \label{DM}\end{equation} where
$\vD(\vq)$ is a real vector that satisfies $\vD(\vq)=-\vD(-\vq)$.
Invariance of $H_{DM}$ under point group operations leads to the
constraint $\tilde{R}\vD_{R\vq}=\vD_{\vq}$, where $\tilde{R}$ is
the proper part of the rotation ($\tilde{R}=Det(R)\times R$). Note
that $\vg(\vsk)$ satisfies the same symmetry relation
($\tilde{R}\vg_{R\vsk}=\vg_{\vsk}$). Consequently, the two vectors
$\vD(\vq)$ and $\vg_{\vq}$ are not orthogonal. Extracting the
pairing contribution from Eq.~\ref{DM} leads to the following
mixed singlet-triplet pairing interaction
\begin{equation}
\left[ -\vD(\vsk-\vsk')\cdot
\hat{\vtau}\right]_{s_{1}s_{2}}\hat{\tau}^{\dag}_{s'_{2}s'_{1}}+\hat{\tau}_{s_{1}s_{2}}\left[\vD(\vsk-\vsk')\cdot
\hat{\vtau}\right]^{\dag}_{s'_{2}s'_{1}}.
\end{equation}
If we assume the form $\vD(\vk-\vk')=e_m(\vg_{\vsk'}-\vg_{\vsk})$
and impose symmetry constraints that arise from Pauli exclusion,
we arrive at the mixed parity term in Eq.~2. As will be shown
below, this form for $\vD$ can be justified.

\subsection{DM Interaction in a weakly interacting Fermi liquid}

The DM interaction is known to exist in materials without
inversion symmetry. This implies that the DM interaction should
arise as a consequence of the existence of $\vg_{\vsk}$ in the
single particle Hamiltonian. Here we calculate the contribution of
$\vg_{\vsk}$ to the DM interaction through a calculation of the
spin susceptibility. In the normal state the static
spin-susceptibility is given by
\begin{eqnarray}
        \chi_{i j}(\vq)&=&-\mu^{2}_{B} k_{B}T \sum_{\vsk} \sum_{\omega_{n}}
        tr \{ \hat{\sigma}_{i} \hat{G}(\vk, \omega_{n})
        \hat{\sigma}_{j} \hat{G}(\vk+\vq, \omega_{n})) \} \nonumber
        \\.
\end{eqnarray}
Upon using the form for the normal state Green's functions and
carrying out the trace over the spins it can be shown
\begin{eqnarray}
        \chi_{i j}(\vq)&=&-2\mu^{2}_{B} k_{B}T \sum_{\vsk} \sum_{\omega_{n}}
        \left\{\left[  \delta_{ij}(1-\hvg_{\vk}\cdot\hvg_{\vk+\vq})\right.\right. \nonumber \\
        &+&(\hve_i\cdot\hvg_{\vk})(\hve_j\cdot\hvg_{\vk+\vq}) + (\hve_i\cdot\hvg_{\vk+\vq})(\hve_j\cdot\hvg_{\vk})  \nonumber\\
        &+& \left. i(\hve_i\times\hve_j)\cdot(\hvg_{\vk+\vq}-\hvg_{\vk})\right]G_1(\vk,\omega_{n})G_1(\vk+\vq,\omega_{n})\nonumber \\
         &+&\left[ \delta_{ij}(1-\hvg_{\vk}\cdot\hvg_{\vk+\vq})+(\hve_i\cdot\hvg_{\vk})(\hve_j\cdot\hvg_{\vk+\vq})      \right. \nonumber \\
         &+& \left. (\hve_i\cdot\hvg_{\vk+\vq})(\hve_j\cdot\hvg_{\vk})-i(\hve_i\times\hve_j)\cdot(\hvg_{\vk+\vq}-\hvg_{\vk})\right] \nonumber \\
        &\cdot & G_2(\vk,\omega_{n})G_2(\vk+\vq,\omega_{n})\nonumber \\
        &+&\left[ \delta_{ij}(1+\hvg_{\vk}\cdot\hvg_{\vk+\vq})-(\hve_i\cdot\hvg_{\vk})(\hve_j\cdot\hvg_{\vk+\vq}) \right. \nonumber \\
        &-& (\hve_i\cdot\hvg_{\vk+\vq})(\hve_j\cdot\hvg_{\vk})-\left. i(\hve_i\times\hve_j)\cdot(\hvg_{\vk+\vq}+\hvg_{\vk})\right] \nonumber\\
        &\cdot& G_1(\vk,\omega_{n})G_2(\vk+\vq,\omega_{n})\nonumber\\
        &+&\left[
        \delta_{ij}(1+\hvg_{\vk}\cdot\hvg_{\vk+\vq})-(\hve_i\cdot\hvg_{\vk})(\hve_j\cdot\hvg_{\vk+\vq}) \right. \nonumber \\
        &-& \left. (\hve_i\cdot\hvg_{\vk+\vq})(\hve_j\cdot\hvg_{\vk}) + i(\hve_i\times\hve_j)\cdot(\hvg_{\vk+\vq}+\hvg_{\vk})\right] \nonumber \\
        &+&\left. G_2(\vk,\omega_{n})G_1(\vk+\vq,\omega_{n})\right\}\nonumber
\end{eqnarray}
The DM interaction is antisymmetric under interchange of $i$ and
$j$:
\begin{eqnarray}
\vD(\vq)&=&-2\mu^{2}_{B} k_{B}T
\sum_{\vsk,\omega_{n}}\left\{\left[G_1(\vk,\omega_{n})G_1(\vk+\vq,\omega_{n}) \right. \right. \nonumber \\
&&- \left. G_2(\vk,\omega_{n})G_2(\vk+\vq,\omega_{n})\right] [\hvg_{\vk+\vq}-\hvg_{\vk}]  \nonumber \\
&&+  \left[ G_2(\vk,\omega_{n})G_1(\vk+\vq,\omega_{n}) \right. \nonumber \\
&&- \left. \left. G_1(\vk,\omega_{n})G_2(\vk+\vq,\omega_{n}) \right] [\hvg_{\vk+\vq}+\hvg_{\vk}] \right\}. \nonumber \\
\end{eqnarray}
Carrying out the sum over Matsubara frequencies, expanding the
expression for $\vD(\vq)$ to linear order in $\vq$ one finds
\begin{eqnarray}
\vD(\vq)&=&2\mu^{2}_{B}\sum_{\vsk}\left\{\left[\frac{\vnabla
n_2\cdot \vq}{\vnabla\xi_2\cdot \vq}-\frac{\vnabla n_1\cdot
\vq}{\vnabla\xi_1\cdot \vq}\right]\vq\cdot\vnabla \hvg  \right. \nonumber \\
&&+ \left. \hvg \vq
\cdot\left[\frac{n_1-n_2}{(\xi_1-\xi_2)^2}(\vnabla \xi_1+\vnabla
\xi_2)-\frac{\vnabla n_1+ \vnabla
n_2}{\xi_1-\xi_2} \right] \right\} \nonumber \\
\end{eqnarray}
where $\xi_i=\xi_i(\vk)$, $n_i=(e^{\beta \xi_i}+1)^{-1}$ is the
Fermi distribution function for band $i$, and
$\vnabla=\vnabla_\vk$. To linear order in $\alpha$, this gives
\begin{equation} 
\vD(\vq)  =-\frac{8\mu_B^2\alpha}{3}\sum_{\vk}\frac{d^2n(\epsilon_{\vk})}{d\epsilon^2_{\vk}}
\vq\cdot\vnabla_{\vk}\vg_{\vsk} .
\end{equation}
For example if $\vg_{\vk}=\vk/k_F$ (valid for a material with
point group $O$), this gives \begin{equation}
\vD(\vq)=-\frac{8\mu_B^2 \alpha}{3} N_0' \frac{\vq}{k_F} = - \frac{4}{3} \chi_P \frac{\alpha N_0'}{N_0} \frac{\vq}{k_F}
\end{equation}
where $N_0'= \frac{dN_0}{d\xi_F}|_{\xi=0}$ is the derivative of the
density of states evaluated at the Fermi surface and $ \chi_P $ the Pauli susceptibility. 
This is a perturbative form and the derivative of density of states is considered at the
Fermi surface in the absence of spin-orbit coupling. Estimates for $ \alpha $ show that
the value can be a considerable fraction of the band width.

\subsection{DM interaction within the Hubbard for finite ASOC }
Here we derive the DM interaction in the strong coupling regime by
finding the effective Hamiltonian that governs the low- energy
excitations of the Hubbard model with ASOC in the large $U/t$
regime. The technique used to extract the Hamiltonian for the
low-energy excitations is similar to that used for the derivation
of the t-J model starting from the Hubbard model.

We choose as zeroth-order Hamiltonian the on site Coulomb
repulsion
\begin{equation}
     {\cal U} = U \sum_i n_{i \uparrow} n_{i \downarrow}
\end{equation}
The eigenstates of ${\cal U}$  are Fock states in the Wannier
representation.  ${\cal U}$ divides the Fock space into two sub
spaces:
\begin{eqnarray}
S&=&\left[ | n_{1\uparrow}, n_{1 \downarrow}, n_{2 \uparrow}, \cdots \rangle :  \forall n_{i \downarrow}+n_{i \uparrow} \leq 1  \right] \nonumber \\
D&=&\left[ | n_{1\uparrow}, n_{1 \downarrow}, n_{2 \uparrow}, \cdots \rangle : \exists n_{i \downarrow}+n_{i \uparrow} = 2  \right]. \nonumber \\
\end{eqnarray}

$D$ contain at least one doubly occupied site, and $S$ are all
configurations with either one or zero electrons per site. The
hopping
\begin{equation}
     {\cal T} = {\cal T}^s + {\cal T}^a
\end{equation}
where
\begin{eqnarray}
     {\cal T}^s &=& - \sum_{i j, s s'}  t_{ij} \sigma_0 c_{i s}^{\dag} c_{j s'} \nonumber \\
     {\cal T}^a &=& - \sum_{i j, s s'} i \valpha_{ij} \cdot \vsig_{s s'} c_{i s}^{\dag} c_{j s'}
\end{eqnarray}
contains now an antisymmetric term ($\valpha_{ij}=-\valpha_{ji}$)
corresponding to the ASOC contribution. This means that also the
effective interaction resulting by the superexchange process has
an antisymmetric part. This last correspond to the DM interaction
\begin{eqnarray}
{\cal H}^{DM}&=& -\sum_{ij, s_1\cdots s_4}  i \frac{t_{ij}  \valpha_{ij} \cdot ( \vsig_{s_1 s_2} \delta_{s_3 s_4}  - \vsig_{s_3 s_4}\delta_{s_1 s_2} )}{U} \nonumber \\
& &c_{i s_1}^{\dag} c_{j s_2} n_{j \uparrow} n_{j \downarrow} c_{j s_3}^{\dag} c_{i s_4} \nonumber \\
&=& \sum_{i j} \frac{4 i t_{i j}}{U} \valpha_{i j} \cdot\left( \vS_i \times \vS_j \right). \nonumber \\
\end{eqnarray}
The interaction ${\cal H}^{DM}$ is a factor $\alpha/t$ smaller
than the usual spin-spin interaction constant $J$. While
$\alpha/t$ will be less than one, it is not necessarily small.

\section{Self-consistent equation}

The self consistent equations (\ref{SelfConsistLin}) requires the
evaluation of $\tomega_n$ and $\vDelta=(\psi,d)$. The
corresponding equations are given by the substitution of Eqs.
(\ref{GreenG}) and (\ref{GreenF}) into Eqs. (\ref{GapEqn1}) and
(\ref{Self}),
\begin{eqnarray}
\label{SelfConsist1}
\tomega_{n} &=& \omega_{n} + \tomega_{n} \Gamma Q_{1}(i \omega_{n}), \\
\label{SelfConsist2}
\tpsi &=& \psi + \tpsi \, \Gamma Q_{1}(i \omega_{n})+ d \, \Gamma  Q_{2}(i \omega_{n}),  \\
\label{SelfConsist3}
    \frac{1}{N_{0}V} \vDelta &=&\pi k_{B} T \sum^{\epsilon_{c}}_{\omega_{n}>-\epsilon_{c}} \hat{Q}(i \omega_{n}) \vDelta, 
\end{eqnarray}
with
\begin{eqnarray}
\hat{Q}(i \omega_n)&=& \frac{Q_{1}(i \omega_{n})}{1-\Gamma Q_{1}(i \omega_{n}) } \left( \begin{array}{cc} -e_{s} & 0  		\\ e_{m} & 0  \end{array} \right)  \nonumber \\
    	         &+& \frac{Q_{2}(i \omega_{n})}{1-\Gamma Q_{1}(i \omega_{n})} \left( \begin{array}{cc} e_{m} & -e_{s}  \\ -e_{t} & e_{m}  \end{array} \right)  \nonumber \\
    &+&   \left\{ Q_3(i \omega_n)+\frac{\Gamma Q^2_2(i \omega_n)}{1-\Gamma Q_{1}(i \omega_{n})} \right \} \left( \begin{array}{cc} 0 & e_{m}  \\
       0 & -e_t
    \end{array} \right)  \nonumber \\
\end{eqnarray}

where we introduced the functions,
\begin{eqnarray}
Q_{1}(i \omega_{n} )&\equiv&  \frac{1}{2} \left \langle \frac{(1+\delta_{N})}{\sqrt{\tomega^{2}_{n}+|\tpsi + d |\vg_{\vk}||^{2}}}+\frac{(1-\delta_{N})}{\sqrt{\tomega^{2}_{n}+|\tpsi - d |\vg_{\vk}| |^{2}}} \right \rangle_{\vsk}, \nonumber \\
Q_{2}(i \omega_{n} )&\equiv& \frac{1}{2} \left \langle \frac{(1+\delta_{N})|\vg_{\vk}|}{\sqrt{\tomega^{2}_{n}+|\tpsi + d |\vg_{\vk}||^{2}}}-\frac{(1-\delta_{N})|\vg_{\vk}|}{\sqrt{\tomega^{2}_{n}+|\tpsi - d |\vg_{\vk}| |^{2}}} \right \rangle_{\vsk} \nonumber, \\
Q_{3}(i \omega_{n} )&\equiv& \frac{1}{2} \left \langle \frac{(1+\delta_{N})|\vg_{\vk}|^2}{\sqrt{\tomega^{2}_{n}+|\tpsi + d |\vg_{\vk}||^{2}}}+\frac{(1-\delta_{N})|\vg_{\vk}|^2}{\sqrt{\tomega^{2}_{n}+|\tpsi - d |\vg_{\vk}| |^{2}}} \right \rangle_{\vsk}.  \nonumber \\
\end{eqnarray}
The parameter $\delta_{N}$
fix the distribution of the DOS at the Fermi level of the two
bands, $N_{1}= N_{0}(1+ \delta_{N})$ and $N_{2}=
N_{0}(1-\delta_{N})$.


\begin{references}
\bibitem{footnote} Note that the in some cases ${\bm g}_{\bm k}$ cannot be attributed to a single irreducible representation. In this case the representation $ \Gamma^-_n $ is composite containing more than one irreducible represenation. We avoid this case here, since our aim is to give a description of pairing in CePt$_3$Si where $ \Gamma^-_n $ is irreducible.
\end{references}
\end{document}